\documentclass[aps,prb,twocolumn,floats,showpacs]{revtex4}
\usepackage{bm}
\usepackage{graphicx}
\begin{document}
%\title{Fermionic Chern-Simons Theory for Fractional Quantum Hall Effect in Graphene}
\title{Fermionic Chern-Simons Theory of SU(4) Fractional Quantum Hall Effect}
\author{Sanhita Modak, Sudhansu S. Mandal, and K. Sengupta}
\affiliation{Department of Theoretical Physics, Indian Association
for the Cultivation of Science, Jadavpur-700032, Kolkata, India}

\begin{abstract}

We develop a Fermionic Chern-Simons (CS) theory for the fractional
quantum Hall effect in monolayer graphene with SU(4) symmetry,
arising from the spin and the valley degrees of freedom, which
involves four distinct CS gauge fields. We choose the corresponding
elements of the CS coupling matrix such that an even number of spin
and valley quantum number dependent flux quanta is attached to all
electrons and that any electron with a given spin and valley quantum
number sees an integer number of flux attached to other electrons
with different (spin and valley) quantum numbers. Using this CS
matrix, we obtain a list of possible fractional quantum Hall states
that might occur in graphene and propose wavefunctions for those
states. Our analysis also applies to fractional quantum Hall states
of both bilayer quantum Hall systems without spin polarization and
bilayer spin polarized graphene.

\end{abstract}
\maketitle

\section{Introduction}

The strong correlation arising out of a complete quench of the
kinetic energy of the electrons in a two dimensional (2D) system in
the presence of a strong perpendicular magnetic field $B$ leads to
the striking phenomenon called fractional quantum Hall effect
(FQHE)\cite{Tsui}. In the fractional quantum Hall regime, the
applied magnetic field is strong enough to make the lowest Landau
level have more states than the number of electrons in the system
leading to a huge degeneracy which is lifted only by the
electron-electron interaction leading to the fractional quantum Hall
states (FQHS) \cite{laughlin1}. A way to understanding the nature of
these states is provided by the composite Fermion(CF) theory
\cite{Jain} in which the state of the system is described in terms
of CFs which correspond to electrons bound to an even number $(2k)$
of vortices of flux quantum $\phi_0=hc/e$. These CFs are weakly
interacting and experience a reduced effective magnetic field
$B^*=B-2k\phi_0\rho$, where $\rho$ is the electron density. The
integer quantum Hall effect (IQHE) of CFs with filling factor
$\nu^*=p$, where $p$ is an integer, can be shown to describe FQHE of
electrons with filling factor $\nu = p/(2kp\pm 1)$.

The CFs posses topological character due to the attached quantized
vortices. This feature of the CF theory can also be understood by
carrying out Chern-Simon (CS) transformation on the electron field
operators.
% $\psi$ leading to a quasiparticle field operator $\psi'$:
%$\psi' = \psi \exp(-i \int d^2 r' {\bm a}(r-r') \rho(r'))$
%\cite{LF1,HLR}.
Such a transformation \cite{LF1,HLR} leads to the introduction of a
topological CS vector potential $\bm{a}$ which produces CS magnetic
field, $b (\bm{r}) \equiv (1/e)\bm{\nabla}\times \bm{a} = 2k\phi_0
\rho(\bm{r})$, proportional to the electron density $\rho(\bm{r})$.
The factor $2k$ ensures that the statistics of the quasiparticles
remain Fermionic. The main difference between the CS quasiparticles
and CFs is that the flux attachment for the former is singular at
the electron's position while the vortex associated with a CF has
finite size and is hence free from any such singularity. To
distinguish CFs from these quasiparticles, we shall refer to them as
CS-CFs in the rest of this work. We note that in spite of the
difference mentioned above, the effective magnetic field seen by
CS-CFs is $B^*$ as in the case of CFs and the IQHE of CS-CFs also
reproduce FQHE of electrons with $\nu = p/(2kp\pm 1)$ in the lowest
Landau level (LLL). The Fermionic CS theory has also been applied to
systems with SU(2) symmetry \cite{MR,LF2}. For example, the FQHS in
LLL of a single layer system with spin degree of freedom \cite{MR}
and a double layer system \cite{LF2} with frozen spin has been
described using such an analysis. It is well-known that the CS
method correctly describes spin (layer) polarizations for FQHS for a
single layer system with spin degree of freedom (bilayer system with
frozen spin).

More recently, both IQHE \cite{IQHE1,IQHE2} and FQHE
\cite{FQHE1,FQHE2,FQHE3} have been observed in single layer graphene
whose effective low-energy theory is described in terms of
Dirac-like quasiparticles \cite{neto1}. These Dirac quasiparticles
are centered around the edges of the hexagonal Brillouin zone of
graphene which hosts the Dirac cones. There are six such cones;
however only two of them turn out be inequivalent since the rest of
the cones differ from one of these by reciprocal lattice vectors.
This leads to two inequivalent species of Dirac quasiparticles in
graphene. These quasiparticles, apart from their physical spin,
carries an additional quantum number which specifies their location
in the graphene Brillouin zone. This is commonly known as the valley
quantum number. The valley degree of freedom thus acts like a
fictitious spin providing an additional internal symmetry to these
quasiparticles. In the presence of physical spin and in the absence
of any symmetry breaking interactions, the internal symmetry of
these quasiparticles is thus SU(4). This is manifested in the Hall
conductivity for IQHE in graphene, $\sigma_{xy} = 4(n+1/2)e^2/h$ for
integer $n$, where the factor $4$ arises due to the spin and the
valley degrees of freedom.

FQHE in the LLL for graphene has been studied by a number of authors
\cite{CF1,TC,CF2, Regnault, Khv}. In Refs.\ \onlinecite{CF1,TC,CF2},
FQHS for spin-polarized electrons, {\it i.e.}, when the FQHS are
SU(2) symmetric due to the valley degrees of freedom, has been
studied. However, given that the Zeeman energy in graphene is small
compared to the Landau level splitting (the ratio of the two is
approximately $10^{-4}$ for $B \sim 1 T$), a full SU(4) symmetric
FQHE seems to be more relevant in graphene. Such SU(4) symmetric
FQHS has been studied using SU(4) generalized CF wave functions
\cite{CF2} and Halperin-like wave functions \cite{Regnault}. The
former \cite{CF2} described a restricted class of filling factor
which arises from equal even integral flux $2k$ attached to each
species of Dirac quasiparticles leading to  $\nu =
(\nu_1+\nu_2+\nu_3+\nu_4)/[2k(\nu_1+\nu_2+\nu_3+\nu_4)\pm 1]$ with
$\nu_1,\nu_2,\nu_3,\nu_4$ being the effective integer filling
factors of four different species of CFs. In this scheme, one
obtains the spin and the valley polarizations of the FQHS for a
given $\nu$ depending on the individual values of $\nu_i$'s (keeping
their sum fixed). For example, the CF scheme describes $\nu = 4/(8k+
1)$ with zero spin and valley polarizations for
$\nu_1=\nu_2=\nu_3=\nu_4=1$ which corresponds to intra-species
exponent $2k+1$ and inter-species exponent $2k$ in the Jastrow form
of the corresponding CF wave functions \cite{CF2}. In contrast, the
work of Ref.\ \onlinecite{Regnault} computes the spin and the valley
polarization directly from the proposed Halperin-like wave
functions. Interestingly, Ref.\ \onlinecite{Regnault} describes some
FQHS which does not have definite spin, valley or mixed
polarizations. These states do not feature in Ref.\
\onlinecite{CF2}. This contradiction, to the best of our knowledge,
has not been yet resolved in the sense that there is no uniform
formalism which reproduces all FQHS obtained from both these
methods. Finally, Ref.\ \onlinecite{Khv} computes Hall conductivity
using a SU(4) symmetric CS action, but does not aim to analyze the
details of the several possible FQHS.

In this paper, we develop a CS theory for SU(4) FQHE which is
relevant for monolayer graphene. The central point of our work is to
introduce a general flux attachment scheme by using a CS coupling
matrix. We choose the corresponding elements of this coupling matrix
such that an even number of flux quanta, which may depend on the
spin and valley quantum numbers, is attached to all electrons and
that any electron with a given spin and valley quantum number sees
an integer number of flux attached to other electrons with different
(spin and valley) quantum numbers. Using this CS matrix, we obtain a
list of possible FQHS that might occur in
graphene and also propose wave function for those states. We show
that our formalism not only reproduces the FQHS obtained in
Ref.\ \onlinecite{CF2,Regnault} and thus resolves the contradiction
mentioned above, but also provides an exhaustive list of other
possible FQHS in this system including those which do not have any
SU(2) analogues. We provide an exhaustive chart of these FQHS for
several filling factors along with various polarizations (valley,
spin and mixed) for each of them. Finally, we note that our analysis
is directly relevant for FQHS in both spin-polarized bilayer
graphene (with layer and valley degrees of freedom providing the
SU(4) symmetry) and conventional bilayer quantum Hall systems (with
layer and spin degrees of freedom providing the SU(4) symmetry).

The organization of the rest of the paper is as follows. In Sec.\
\ref{formalsim}, we develop the SU(4) CS theory, derive equations
for the filling factor $\nu$, the spin (S), the valley (V) and the
mixed (M) polarizations using this theory, and propose wavefunctions
which describes the obtained FQHS. This is followed by Sec.\
\ref{analysis}, where we analyze these equations to provide an
exhaustive list of possible FQHS for monolayer graphene. Finally, we
summarize our results and conclude in Sec.\ \ref{conc}.

\section{Chern-Simons Formalism}
\label{formalsim}

The low-energy states in graphene can be described by an effective
Dirac-like Hamiltonian
\begin{eqnarray}
 {\cal H} &=& \int d\bm{r} \psi_e^\dagger (\bm{r}) H \psi_e (\bm{r}) \nonumber \\
 & +& \frac{1}{2}\int d\bm{r}\int d\bm{r'}V(\bm{r}-\bm{r'})
 :\hat{\rho}_e(\bm{r})\hat{\rho}_e(\bm{r}'): \label{ham1},
\end{eqnarray}
where $\psi_e$ is the eight component electronic annihilation
operator whose components correspond to the sublattice, the valley
and the spin degrees of freedom \cite{neto1}, $\hat{\rho}_e =
\psi_e^\dagger \psi_e$ is the density operator, $:..:$ denotes
normal ordering, $V(\bf r)$ represents electron-electron interaction
whose precise form is unimportant for our purpose, and
\begin{equation}
 H = v_F\left( \begin{array}{cccc}
            \bm{\sigma}\cdot \bm{\Pi} &  0 & 0 & 0\\
 0 & \bm{\sigma}\cdot \bm{\Pi} &  0 & 0\\
            0& 0 & (\bm{\sigma}\cdot \bm{\Pi})^T & 0 \\
0& 0 & 0 & (\bm{\sigma}\cdot \bm{\Pi})^T
\end{array} \right),
\end{equation}
with $\bm{\Pi} = -i \bm{\nabla} +e\bm{A}$ and $\bm{\nabla} \times
\bm{A} = B \hat{z}$. Here and in the rest of this work, we shall set
$\hbar = c=1$. Here $\sigma$'s are Pauli matrices which describe two
sublattice in graphene. In the rest of this work, we shall use the
shorthand notation $1\equiv (\uparrow\, ,\,+)$, $2\equiv
(\downarrow\, ,\, +)$, $3\equiv (\uparrow\,,\,-)$, and $4\equiv
(\downarrow\, ,\, -)$, where $\uparrow ,\downarrow$ represent the
physical spin state and $\pm$ (where $+$ corresponds to an electron
in the $K$ valley) represent the valley states of the graphene
electron.

 \begin{figure}
\includegraphics[width=3.5in,height=3.5in]{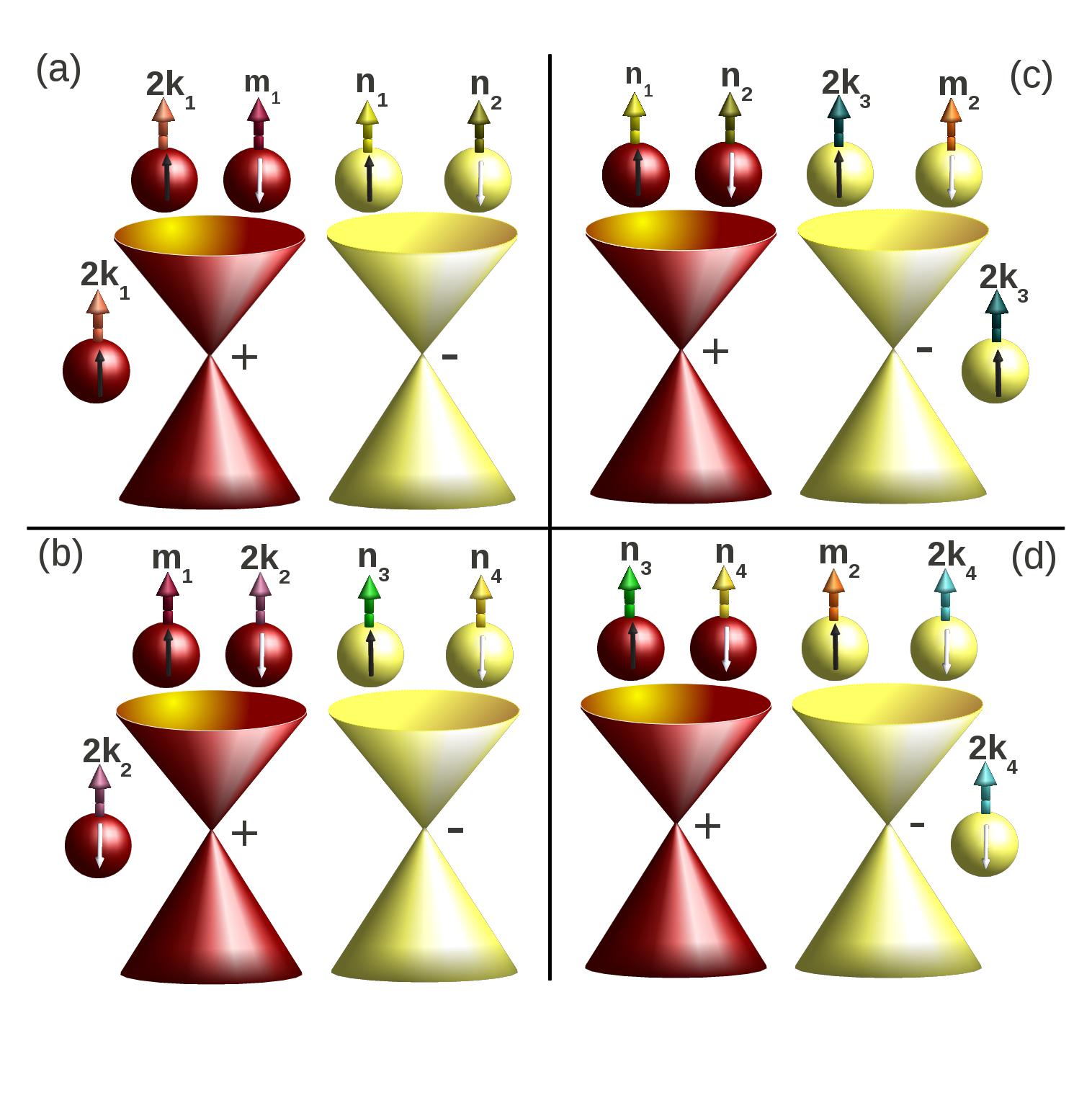}
%\rotatebox{0}{
%\includegraphics*[bb=0 0 246 238, width=\linewidth]{Fignew.jpg}}
\caption{(color online) A pictorial representation of the flux
attachment scheme. The two distinct Dirac cones in a single layer
graphene are represented as $\pm$ valleys. The spheres represent
electrons and their colors (identical to that of the Dirac cones)
denote the valleys in which the electrons belong. The arrows
pointing up(down) on the surface of the spheres represent the spins
of the electrons. The thick arrows represent the flux attached to
each electron. The number of flux quanta attached to each CS-CF with
respective spin and valley are (a) $2k_1$ for $(\uparrow,+)$, (b)
$2k_2$ for $(\downarrow,+)$, (c) $2k_3$ for $(\uparrow,-)$, and (d)
$2k_4$ for $(\downarrow,-)$. The numbers written above the flux
lines in each of the panels represent the number of flux quanta
attached to the four different CS-CFs as seen by a (a)
$(\uparrow,+)$, (b) $(\downarrow,+)$, (c) $(\uparrow,-)$, and (d)
$(\downarrow,-)$ CS-CF.} \label{Figure}
\end{figure}

We now introduce CS-CF quasiparticle creation operator
$\psi^\dagger(\bm{r})$ in terms of $\psi_e^\dagger(\bm{r})$:
\begin{equation}
 \psi^\dagger_{e,\alpha}(\bm{r}) \rightarrow \psi_\alpha^\dagger
 (\bm{r})\exp\left[-i{\cal K}_{\alpha\beta}\int d\bm{r'}\rm{arg}(\bm{r}-\bm{r'})
\rho_\beta(\bm{r'})\right],
\end{equation}
where the indices $\alpha$ and $\beta$ takes value from $1$ to $4$
described above and thus represents 4-components of the Dirac spinor
in the spin and the valley space. Note that the CS transformation is
a scalar in the pseudospin space and thus do not mix the sublattice
indices of the electrons. Here $\rm{arg}(\bm{r}-\bm{r}')$ represents
the angle made by the vector $(\bm{r}-\bm{r}')$ with x-axis and the
explicit form of the matrix $\cal{K}$ is given by
\begin{equation}
\cal{K} = \left(
      \begin{array}{llll}
         2k_1 & m_1 &n_1 & n_2 \\
         m_1 & 2k_2 & n_3 & n_4 \\
         n_1 & n_2  & 2k_3 & m_2 \\
         n_3 & n_4 & m_2 & 2k_4
      \end{array}
\right), \label{K-matrix}
\end{equation}
where $k$'s, $m$'s and $n$'s are positive integers including zero.
Note that the number of flux quanta attached to a CS-CF of species
$\beta$ as seen by CS-CFs of species $\alpha$ is given by the
component ${\cal K}_{\alpha\beta}$ (Eq.~(\ref{K-matrix})). This is
schematically shown in Fig.\ \ref{Figure}. The diagonal elements of
${\cal K}$ are chosen to be even integers so as to ensure the
Fermionic statistics of the CS-CF quasiparticles.

Such a transformation leads to the effective Hamiltonian for the CS
quasiparticles
\begin{eqnarray}
 {\cal H}_{\rm{qp}} &=& \int d \bm{r} \psi_\alpha^\dagger (\bm{r})
 H_{\rm{eff}} \psi_\alpha (\bm{r}) \nonumber \\
&+& \frac{1}{2}\int d \bm{r}\int d \bm{r}'
V(\bm{r}-\bm{r'}):\hat{\rho}(\bm{r})\rho(\bm{r}'):,
\end{eqnarray}
with
\begin{equation}
 H_{\rm{eff}} = v_F\left( \begin{array}{cccc}
            \bm{\sigma}\cdot \bm{\tilde{\Pi}}_1 &  0 &0 & 0\\
0 & \bm{\sigma}\cdot \bm{\tilde{\Pi}}_2 &  0 &0 \\
             0 & 0 & (\bm{\sigma}\cdot \bm{\tilde{\Pi}}_3)^T & 0 \\
0 & 0 & 0 & (\bm{\sigma}\cdot \bm{\tilde{\Pi}}_4)^T
 \end{array} \right),
\end{equation}
where $\bm{\tilde{\Pi}}_\alpha = -i \bm{\nabla}
+e\bm{A}-\bm{a}_\alpha$. The 4-component CS gauge fields ${\bm
a}_{\alpha}$ so obtained are given by
\begin{eqnarray}
\bm{a}_\alpha = {\cal K}_{\alpha\beta}\int d
\bm{r}'g(\bm{r}-\bm{r'})\hat{\rho}_\beta(\bm{r'}), \label{cspot}
\end{eqnarray}
where $g(r) = (\hat{z}\times \bm{r})/r^2$. The corresponding
4-component CS magnetic fields are then given by
\begin{equation}
b_\alpha\equiv (1/e)\bm{\nabla}\times \bm{a}_\alpha  = \phi_0{\cal
K}_{\alpha\beta}\,\,\rho_\beta (\bm{r}). \label{csfield}
\end{equation}
Eq.\ \ref{csfield} shows that each of the species of the CS-CF will
experience different mean effective magnetic fields
$B^{\ast}_{\alpha}$. The relation between these mean effective
fields and the total applied physical field $B$ is given by
\begin{equation}
 B^*_\alpha = B - \phi_0 \cal{K}_{\alpha\beta}\,\,\rho_{\beta}.
\end{equation}
where $\rho$'s are the mean densities.
Note that $B^*_\alpha$ creates different set of effective Landau
levels for the CS-CFs akin to Landau levels for electrons due to
$B$. Denoting $\nu_1$, $\nu_2$, $\nu_3$, and $\nu_4$ be the number
of effective Landau levels completely filled by the respective
species of CS-CFs, one obtains the relation \cite{MR}
\begin{equation}
\frac{\rho_\alpha}{\nu_\alpha} = \frac{\rho}{\nu} -
\cal{K}_{\alpha\beta}\,\,\rho_{\beta}. \label{meanfield}
\end{equation}
By defining the spin ($S$), the valley ($V$), and the mixed ($M$)
polarizations respectively as
\begin{eqnarray}
S &=& (\rho_1+\rho_3-\rho_2-\rho_4)/\rho,\nonumber\\
V &=& (\rho_1+\rho_2-\rho_3-\rho_4)/\rho, \nonumber\\
M &=& (\rho_1+\rho_4-\rho_2-\rho_3)/\rho, \label{pol1}
\end{eqnarray}
Eq.~(\ref{meanfield}) can be expressed as
\begin{eqnarray}
 & &(1+S+V+M)(2k_1+\frac{1}{\nu_1}) + (1+V-S-M)m_1 \nonumber \\
& & +(1-V)(n_1+n_2) +(S-M)(n_1-n_2) = \frac{4}{\nu} \label{eq1} ,\\
& & (1-S+V-M)(2k_2+\frac{1}{\nu_2}) + (1+V+S+M)m_1 \nonumber \\
 & &+(1-V)(n_3+n_4) +(S-M)(n_3-n_4) =  \frac{4}{\nu} \label{eq2} ,\\
& & (1+S-V-M)(2k_3+\frac{1}{\nu_3}) + (1-V-S+M)m_2  \nonumber \\
& &+(1+V)(n_1+n_2) +(S+M)(n_1-n_2) =  \frac{4}{\nu} \label{eq3} ,\\
& & (1-S-V+M)(2k_4+\frac{1}{\nu_4}) + (1-V+S-M)m_2  \nonumber \\
& &+(1+V)(n_3+n_4) +(S+M)(n_3-n_4) = \frac{4}{\nu} \label{eq4} .
\end{eqnarray}
Eqs.~(\ref{eq1})..(\ref{eq4}) represent the central result of this
work and provide a relation between the total filling factor $\nu$
and the spin ($S$), the valley ($V$), and the mixed ($M$)
polarizations of a FQHS in terms of the attached flux numbers $k$'s,
$m$'s and $n$'s. The CS theory does not predict these attached flux
numbers; however, one can use them as parameters whose variation
leads to a list of possible FQHS at the saddle point level for a
given $\nu$ with different values of $S$, $V$ and $M$. Thus the CS
method do not predict a definite FQHS for a given filling $\nu$; it
provides a list of possible FQHS. Moreover, in terms of these flux
attachment numbers, one can write down, via a straightforward
generalization of methods used in Refs.\ \onlinecite{MR,LF2}, a
variational wavefunction of these FQHS for the filling factor $\nu$
as
\begin{eqnarray}
&&\Psi (u,v,w,z) = {\cal P}_L \Phi_{\nu_1}(u_1,\cdots,u_{N_1}) \Phi_{\nu_2}(v_1,\cdots,v_{N_2})  \nonumber \\
&&\times \Phi_{\nu_3}(w_1,\cdots,w_{N_3}) \Phi_{\nu_4}(z_1,\cdots,z_{N_4}) \prod_{i<j}^{N_1} (u_i-u_j)^{2k_1}
\nonumber \\
&&\times \prod_{i<j}^{N_2} (v_i-v_j)^{2k_2} \prod_{i<j}^{N_3} (w_i-w_j)^{2k_3} \prod_{i<j}^{N_4} (z_i-z_j)^{2k_4}
\nonumber \\
&&\times\prod_{i,j}^{N_1,N_2} (u_i-v_j)^{m_1}  \prod_{i,j}^{N_3,N_4} (w_i-z_j)^{m_2} \prod_{i,j}^{N_1,N_3} (u_i-w_j)^{n_1}  \nonumber \\
&&\times \prod_{i,j}^{N_1,N_4} (u_i-z_j)^{n_2} \prod_{i,j}^{N_2,N_3}
(v_i-w_j)^{n_3} \prod_{i,j}^{N_2,N_4} (v_i-z_j)^{n_4} \nonumber\\
\label{Wavefn}
\end{eqnarray}
where $u$'s, $v$'s, $w$'s and $z$'s are complex coordinates for the
particles of species $1$, $2$, $3$, and $4$ respectively
\cite{laughlin1}, $N_{\alpha}$ denotes the number of CS-CFs of
species $\alpha$, $\Phi_{\nu_\alpha}$ represents the IQHE
wavefunction for $\nu_\alpha$ filled Landau levels of CS-CF
belonging to species $\alpha$, and ${\cal P}_L$ represents
projection into the LLL. We note here that the CS theory alone
cannot lead to the wavefunction given in Eq.\ (\ref{Wavefn}).
Whereas the CS formalism does provide the exponents of both the
inter- and intra-species Jastrow factors in Eq.\ (\ref{Wavefn}),
both the IQHE wavefunctions and the projection into the LLL receive
input from the CF theory \cite{Jain}. We expect the computation of
the interaction energy of the system using Eq.\ (\ref{Wavefn}) to
provide further information about the FQHS in graphene. However, in
this paper, instead of undertaking such a calculation, we shall be
content to classify the possible FQHS based on Eqs.\
(\ref{eq1})..(\ref{eq4}). We carry out this analysis in the next
section.

\section{Filling factors and polarizations}
\label{analysis}

In this section, we are going to address FQHS corresponding to
specific value of the flux attachment parameters
$\{k_1,k_2,k_3,k_4\}$ and $\{m_1,m_2,n_1,n_2,n_3,n_4\}$, and
effective integer filling factors $\{\nu_1,\nu_2,\nu_3,\nu_4\}$.
While Eqs.~(\ref{eq1})..(\ref{eq4}) may be used to determine $\nu$,
$S$, $V$ and $M$ for any set of parameters (except $M$, $S$ or $V$
may remain undetermined for some special cases which we shall
discuss below), in the rest of this work, we focus on some specific
choices of these parameters which allow simple analytical solution
to these equations. More specifically, we choose the flux seen by
inter-valley CS-CFs and the flux attached to each CS-CF to be
independent of their spin. This imposes the restriction
$n_1=n_2=n_3=n_4=n$ and $k_1=k_2,\, k_3=k_4$. Within this restricted
parameter space, we find analytical solutions to Eqs.\
(\ref{eq1})..(\ref{eq4}) and discuss several possible FQHS which
arise out of these solutions. We arrange these
solutions into several distinct groups below. \\

I. {\bf {\em Arbitrary values of $\nu_1, \nu_2, \nu_3$ and $\nu_4$ }}: \\

(a) $2k_1=2k_3=m_1=m_2=n=2k$: In this case, all the matrix elements
of ${\cal K}$ have the same value since the number of flux quanta
attached to each CS-CFs and as seen by other CS-CFs, irrespective of
the species of CS-CFs, are $2k$. The eigenvalues of ${\cal K}$
matrix for this special case are $8k$ and $0$ (triply degenerate).
Therefore there is only one dynamical CS gauge field ${\cal A}^\mu =
a^\mu_1+a^{\mu}_2+a^\mu_3+a^\mu_4$. The rest of the CS gauge fields
decouple \cite{MR}. Solving Eq.\ (\ref{eq1})..(\ref{eq4}), we find
\begin{eqnarray}
\nu &=& \frac{\nu^*}{2k\nu^*+1}\, ,\, V=\frac{2(\nu_1+\nu_2)-\nu^*}{\nu^*} ,\nonumber \\
S &=& \frac{2(\nu_1+\nu_3)-\nu^*}{\nu^*}\, ,\,
M=\frac{2(\nu_1+\nu_4)-\nu^*}{\nu^*} ,\label{sol1}
\end{eqnarray}
where $\nu^*=\nu_1+\nu_2+\nu_3+\nu_4$. This is precisely the
Toke-Jain sequence in graphene \cite{CF2} and the wavefunction [Eq.\
(\ref{Wavefn})] obtained for these set of parameters exactly matches
with the CF wave function \cite{CF2}. The sequence of FQHS generated
with this set of parameters is same as the SU(2) sequence for
$\nu^*\leq 2$. We note that these states can be degenerate. For
example, for $\nu^*=2$, the two effective Landau levels may be
filled by two different species of CS-CFs, leading to a six-fold
degeneracy of the ground state. We find that in the limit $\nu^* \to
\infty$, $\nu = 1/2k$ and hence these even denominator states
correspond to Fermi sea of CFs \cite{HLR}.

We note that the class of solutions [Eq.\ (\ref{sol1})] which has
equal exponents in the Jastrow factors in their wavefunction as can
be seen from Eq.\ (\ref{Wavefn}). However, the CF wave function for
bilayers, constructed in Ref.\ \onlinecite{Scarola}, do not
necessarily have equal exponents in Jastrow factors for intra- and
inter-layer CFs. Taking cue from this, we now construct solutions to
Eqs.\ (\ref{eq1})..(\ref{eq4}) where the off-diagonal elements of
${\cal K}$ are different from the diagonal elements. We discuss
these solutions as cases (b), and (c) below.\\

(b) $2k_1=2k_3=m_1=m_2=2k$ and $n\neq 2k$: For this choice of
parameters, we find
\begin{eqnarray}
\nu &=& \frac{\nu^*+2(2k-n)(\nu_1+\nu_2)(\nu_3+\nu_4)}{2k\nu^*+1+(4k^2-n^2)(\nu_1+\nu_2)(\nu_3+\nu_4)} ,\nonumber\\
V &=& \frac{2(\nu_1+\nu_2)-\nu^*}{\nu^*+2(2k-n)(\nu_1+\nu_2)(\nu_3+\nu_4)} ,\nonumber \\
S &=& \frac{2(\nu_1+\nu_3)-\nu^*+2(2k-n)(\nu_1\nu_3-\nu_2\nu_4)}{\nu^*+2(2k-n)(\nu_1+\nu_2)(\nu_3+\nu_4)} ,\nonumber \\
M &=&
\frac{2(\nu_1+\nu_4)-\nu^*+2(2k-n)(\nu_1\nu_4-\nu_2\nu_3)}{\nu^*+2(2k-n)(\nu_1+\nu_2)(\nu_3+\nu_4)}
\label{sol2} .
\end{eqnarray}
Note that for these solutions, the even denominator FQHS with $\nu
=1/2k$ occurs when $(2k-n)^2=1/[(\nu_1+\nu_2)(\nu_3+\nu_4)]$. Since
$k$, $n$, and $\nu$'s are integers, the only possible way to satisfy
this condition is to have $(\nu_1+\nu_2)=1=(\nu_3+\nu_4)$ and
$n=2k\pm1$. We point out that the even-denominator FQHS so obtained
do not correspond to $\nu^{\ast} \to \infty$ and hence do not lead
to the formation of a Fermi sea of CS-CF. These states are similar
to those obtained for bilayer
SU(2) quantum Hall systems. \\

(c) $2k_1=2k_3=2k$, $m_1=m_2=m\neq 2k$, and $n\neq 2k$: In this case,
we find solutions with zero spin and mixed polarization ($M=S=0$)
and with finite non-zero valley polarizations given by
\begin{eqnarray}
\nu &=&
\left[4\nu^*+4(2k+m-2n)(\nu_1+\nu_2)(\nu_3+\nu_4)\right]/{\mathcal
D},\nonumber\\
{\mathcal D} &=& 4+2(2k+m)\nu^*
+((2k+m)^2-4n^2) \nonumber\\
&& \times (\nu_1+\nu_2)(\nu_3+\nu_4) ,\nonumber \\
V &=&
\frac{2(\nu_1+\nu_2)-\nu^*}{2\nu^*+2(2k+m-2n)(\nu_1+\nu_2)(\nu_3+\nu_4)}
 \label{sol3} .
\end{eqnarray}
We note that these states do not have any analogue in U(1) and SU(2)
FQHE and can only occur for SU(4) symmetric FQHE. Similarly,  a set
of FQHS can be found, again for $2k_1=2k_3=2k$, $m_1=m_2=m$, and
$n\neq 2k$, where the mixed and the valley polarizations vanish
($M=V=0$), but the spin polarization $S$ remain finite. These states
correspond to
\begin{eqnarray}
\nu &=& \frac{4(\nu_2\nu_3-\nu_1\nu_4)}{\nu^*-2(\nu_1+\nu_4)+(2k+m+2n)(\nu_2\nu_3-\nu_1\nu_4)}, \nonumber \\
S &=& \frac{\nu^*-2(\nu_1+\nu_2)}{(2k-m)(\nu_2\nu_3-\nu_1\nu_4)}.
\label{sol4}
\end{eqnarray}
The same choice of parameters also allows for FQHS with $S=V=0$ and
finite $M$ which are given by
\begin{eqnarray}
 \nu &=& \frac{4(\nu_1\nu_3-\nu_2\nu_4)}{\nu^*-2(\nu_2+\nu_4)+(2k+m+2n)(\nu_1\nu_3-\nu_2\nu_4)}, \nonumber \\
M &=& \frac{2(\nu_1+\nu_2)-\nu^*}{(2k-m)(\nu_1\nu_3-\nu_2\nu_4)}.
\label{sol5}
\end{eqnarray}
Note that none of the above-mentioned states have any analogue in U(1) and SU(2) FQHS.\\

II. {\bf {\em SU(4) singlet case: $\nu_1=\nu_2=\nu_3=\nu_4=1$}} \\

For the SU(4) singlet states which correspond to all $\nu_i=1$, one
has several possible solutions. The filling factors for these class of
solutions which correspond to zero spin, valley and mixed polarizations
($M=S=V=0$)  are given by
\begin{equation}
 \nu = \frac{8}{2+2k_1+2k_3+m_1+m_2+4n} \label{sol6}
\end{equation}
provided that $2k_1+1 \neq m_1$ and $2k_3+1\neq m_2$. Below, we
classify the other SU(4) singlet FQHS. \\

The first such set of states that we classify consists of $M=S=0$
but $V\neq 0$. These FQHS correspond to
 \begin{eqnarray}
\nu &=&
\frac{4(1+k_1+k_3)+2(m_1+m_2-4n)}{(2k_1+1+m_1)(2k_3+1+m_2)-4n^2}
\label{sol7a}, \\
V &=& \frac{2(k_3-k_1)+m_2-m_1}{2(1+k_1+k_3)+(m_1+m_2-4n)},
\label{sol7b}
\end{eqnarray}
and have finite valley but no spin and mixed polarization.

The second class of states correspond to $k_1=k_3=k$. First, we note
that for these states, if we choose $2k+1=m_1=m_2=m$, we find that
     \begin{equation}
\nu = \frac{2}{m+n} \label{sol8},
     \end{equation}
For these states, $M$ and $S$ are undetermined. $V$ can be
determined only if $m\neq n$ ( which correspond to $V$=0); for
$m=n$, $V$ is also undetermined. The filling factors for these
states with $V=0$ are $2/3$, $1/2$, and $2/5$ for $k=1$ and $2/5$,
$1/3$, $2/7$, $1/4$, and $2/9$ for $k=2$. In contrast, the states
for which $V$ is also undetermined have filling factor $1/3$ with
$k=1$ and $1/5$ with $k=2$. The wavefunctions [Eq.\ (\ref{Wavefn})]
for the filling factors $2/3$, $2/5$, and $1/3$ obtained from our
formalism are precisely the same Halperin-like wavefunctions
proposed in Ref.\ \onlinecite{Regnault}. We note that the even
denominator states in the above-mentioned sequence (such as $1/2$
and $1/4$) do not correspond to the Fermi sea of CFs. Second, if
$2k+1=m_1\neq m_2$, the solutions of Eqs.\ (\ref{eq1})..(\ref{eq4})
yields FQHS with $M=S$ but undetermined, and with the values of the
filling fractions and the valley polarizations given by
\begin{equation}
\nu= \frac{3m_1+m_2-4n}{m_1^2+m_1m_2-2n^2}\, ,
V=\frac{m_2-m_1}{3m_1+m_2-4n}. \label{sol9}
\end{equation}
These states do not appear in the work of Ref.\
\onlinecite{Regnault}. Similar states with undetermined $M$ and $S$
and with $M=S$ also occurs for $k_1 \neq k_3$, and
$2k_1+1=m_1=m_2=m$. The corresponding filling fractions and valley
polarizations are given by
\begin{equation}
\nu= \frac{2k_3+1+3m-4n}{(2k_3+1+m)m-2n^2}, V=
\frac{2k_3+1-m}{2k_3+1+3m-4n}. \label{sol10}
\end{equation}
Finally, we note that if one chooses $2k_1+1=m_1$ and $2k_3+1=m_2$,
one finds FQHS with $M$ and $S$ undetermined but not necessarily
equal which yields
\begin{equation}
\nu = \frac{m_1+m_2-2n}{m_1m_2-n^2}\, , V=
\frac{m_2-m_1}{m_1+m_2-2n}
 \,\label{sol11} .
\end{equation}

The filling factors for all the SU(4) singlet FQHS and their
corresponding spin, valley and mixed polarizations are tabulated in
Tables~\ref{table1} and \ref{table2}. We point out that these states
constitutes the simplest possible FQHS in monolayer graphene which
do not have any analogue in U(1) and SU(2) FQHE. We note that the
same states would show up in spin polarized bilayer graphene when
$S$ is interpreted as layer polarization and in quantum Hall bilayer
system where $V$ is interpreted as the layer polarization. Further
the filling factors for some of these FQHS, such as $8/19$, are only
found for SU(4) case \cite{Regnault}. It is interesting to observe
that FQHS with filling factor $2/3$ and $3/5$ can be obtained only
by attaching antiparallel flux to the CFs in CF theory
\cite{Jain,CF1,CF2}. In contrast, the CS analysis suggests these
states may also arise due to parallel flux attachment as seen from
Tables~\ref{table1} and \ref{table2} . Also, while FQHS with filling
factor $\nu = 4/11$ correspond to FQHE of CFs in U(1) or SU(2) CF
theory\cite{MJ,CMJ}, it may arise due to IQHE of CS-CFs in SU(4)
FQHE. Finally, we note that the general formulae provided in Eqs.\
(\ref{eq1})..(\ref {eq4}) may contain many other FQHS. A detailed
numerical analysis of these states is left for future study.

\section{Conclusion}
\label{conc}

In summary, we have developed a Fermionic CS theory for SU(4) FQHE
and analyzed the possible FQHS obtained from such a theory. We have
reproduced SU(4) FQHS arising from CF theory \cite{CF2} as well as
Halperin-like \cite{Regnault} states within a single unified
formalism. We have also proposed several other states which are not
obtained in the previous studies. Although the filling factors and
their polarizations presented here are for monolayer graphene, the
analysis is valid for any SU(4) system. Two other examples of such
systems where this theory could be applicable are bilayer quantum
Hall systems and bilayer graphene \cite{Nomura} with complete spin
or valley polarizations. Taking cue from the CS theory \cite{Jain},
we have proposed wavefunctions for all of these FQHS. We note that
for FQHS with particular filling factor, the precise ground state
wavefunction will depend on the exact nature of the interaction
between electrons. It will be interesting to obtain the overlap of
the ground state with our proposed wavefunction. Finally, the ground
state for FQHS in graphene may be tuned by tuning either the Zeeman
coupling or the  inter-valley coupling. It will certainly be
interesting to use our proposed wavefunction to study the resulting
transitions between the FQHS for all of these states by changing
Zeeman coupling and obtain the corresponding phase diagram. We leave
these issues for future studies.

KS thanks DST for support through grant SR/S2/CMP-001/2009.

\begin{table}
\caption{A chart of the possible filling fractions $\nu$ with
numerator $<5$ for the SU(4) singlet states and the corresponding
polarizations $S$, $V$ and $M$ for different sets of parameters
$\{k_1,k_3,m_1,m_2,n\}$. Here we have chosen $k_1=k_2$, $k_3=k_4$,
$\nu_1=\nu_2=\nu_3=\nu_4=1$, $k_3=1,2$, $k_1 \leq k_3$, $m_1 \leq
2k_1+1$, $m_2 \leq 2k_3+1$, and $n= 0$.. max($m_1 ,m_2$). The symbol
`--' for the polarizations denotes undetermined value. In the last
column, we tabulate the Eq. number from which the corresponding
state has been computed.} \label{table1}
 \begin{tabular}{|c|c|c|c|c|c|c|c|c|c|}\hline
$\,\,\nu\,\,$& $\,\,k_1\,\,$& $ \,\,k_3\,\,$& $\,\,m_1\,\,$ &
$\,\,m_2\,\,$ &$ \,\,n\,\,$ & $\,\,S\,\,$ &$\,\, V\,\,$&$ \,\,M\,\,$
& Eq. No.\\ \hline\hline $1/2$ & 1 & 1 & 3 & 3 & 1 & -- & 0 & -- &
(\ref{sol8})\\ \hline $1/2$  & 1 & 2 & 2 & 3 & 1 & 0 & 1/3 & 0 & (\ref{sol7a}-\ref{sol7b}) \\
\hline $1/2$ & 2 & 2 & 0 & 3 & 1 & 0 & 1/3 & 0 & (\ref{sol7a}-\ref{sol7b})\\ \hline $1/2$
& 2 & 2 & 1 & 1 & 1 & 0 & 0 & 0 & (\ref{sol6})\\ \hline $1/2$ & 2 & 2 & 3 &
3 & 0 & 0 & 0 & 0 &  (\ref{sol6}) \\ \hline \hline $1/3$& 1 & 1 & 3 & 3 & 3 &
-- & -- & -- & (\ref{sol8}) \\ \hline $1/3$& 2 & 2 & 5 & 5 & 1 & -- & 0 &
-- & (\ref{sol8}) \\ \hline\hline $1/4$ &2 &2 & 5 & 5 & 3 &--& 0 &--&  (\ref{sol8})
\\ \hline \hline $1/5$ & 2 & 2 & 5 & 5 & 5 & -- & -- & -- & (\ref{sol8}) \\
\hline \hline $2/3$ & 1 & 1 & 1 & 1 & 1 & 0 & 0 & 0 & (\ref{sol6})\\ \hline
$2/3$ & 1 & 1 & 3 & 3 & 0 & -- & 0 & -- & (\ref{sol8}) \\ \hline $2/3$ & 2 &
2 & 1 & 1 & 0 & 0 & 0 & 0 &  (\ref{sol6}) \\ \hline\hline $2/5$ & 1 & 1 & 3 &
3 & 2 & -- & 0 & -- & (\ref{sol8}) \\ \hline $2/5$ & 2 & 2 & 5 & 5 & 0 & --
& 0 & -- & (\ref{sol8}) \\ \hline $2/5$ & 2 & 2 & 3 & 3 & 1 & 0 & 0 & 0 &
 (\ref{sol6}) \\ \hline \hline $2/7$ & 2 & 2 & 3 & 3 & 3 & 0 & 0 & 0 & (\ref{sol6}) \\
\hline $2/7$ & 2 & 2 & 5 & 5 & 2 & -- & 0 & -- &    (\ref{sol6}) \\ \hline
\hline $2/9$ & 2 & 2 & 5 & 5 & 4 & -- & 0 & -- & (\ref{sol8}) \\ \hline
\hline $3/4$ & 1 & 1 & 0 & 1 & 1 & 0 & 1/3 & 0 & (\ref{sol7a}-\ref{sol7b}) \\ \hline
$3/4$ & 1 & 2 & 1 & 3 & 0 & 0 & 1/3 & 0 &(\ref{sol7a}-\ref{sol7b}) \\ \hline \hline
$3/5$ & 1 & 2 & 1 & 1 & 1 & 0 & 1/3 & 0 & (\ref{sol7a}-\ref{sol7b}) \\ \hline \hline
$3/7$ & 1 & 2 & 2 & 1 & 2 & 0 & 1/3 & 0 &  (\ref{sol7a}-\ref{sol7b})\\ \hline
$3/7$ & 1 & 2 & 3 & 5 & 1 & -- & 1/3 & -- &  (\ref{sol11})\\ \hline\hline
$3/8$ & 1& 2 & 3 & 3 & 2 & -- & 1/3 & -- &  (\ref{sol10}) \\ \hline $3/8$ & 2 & 2 & 1 &
3 & 2 & 0 & 1/3 & 0 & (\ref{sol7a}-\ref{sol7b}) \\ \hline \hline $4/5$ & 1 & 1 & 2 & 2 &
0 & 0 & 0 & 0 &  (\ref{sol6}) \\ \hline $4/5$ & 1 & 2 & 2 & 0 & 0 & 0 & 0 & 0
&  (\ref{sol6}) \\ \hline $4/5$ & 2 & 2 & 0 & 0 & 0 & 0 & 0 & 0 &  (\ref{sol6}) \\
\hline\hline $4/7$ & 1 & 1 & 2 & 2 & 1 & 0 & 0 & 0 &  (\ref{sol6}) \\ \hline
$4/7$ & 1 & 2 & 1 & 3 & 1 & 0 & 1/2 & 0 & (\ref{sol7a}-\ref{sol7b}) \\ \hline $4/7$ & 1 &
2 & 2 & 0 & 1 & 0 & 0 & 0 &  (\ref{sol6}) \\ \hline $4/7$ & 2 & 2 & 2 & 2 & 0
& 0 & 0 & 0 &  (\ref{sol6}) \\ \hline \hline $4/9$ & 1 & 1 & 2 & 2 & 2 & 0 &
0 & 0 &  (\ref{sol6}) \\ \hline $4/9$ & 1 & 2 & 2 & 0 & 2 & 0 & 0 & 0 &  (\ref{sol6})
\\ \hline $4/9$ & 2 & 2 & 2 & 2 & 1 & 0 & 0 & 0 &  (\ref{sol6}) \\ \hline
$4/9$ & 2 & 2 & 4 & 4 & 0 & 0 & 0 & 0 &  (\ref{sol6}) \\ \hline \hline $4/11$
& 1 & 2 & 3 & 5 & 2 & -- & 1/2 & -- &  (\ref{sol11}) \\ \hline $4/11$ & 2 & 2
& 2 & 2 & 2 & 0 & 0 & 0 &  (\ref{sol6}) \\ \hline $4/11$ & 2 & 2 & 4 & 4 & 1
& 0 & 0 & 0 &  (\ref{sol6}) \\ \hline \hline $4/13$ & 2 & 2 & 4 & 4 & 2 & 0 &
0 & 0 &  (\ref{sol6}) \\ \hline \hline $4/15$ & 2 & 2 & 4 & 4 & 3 & 0 & 0 & 0
&   (\ref{sol6}) \\ \hline \hline $4/17$ & 2 & 2 & 4 & 4 & 4 & 0 & 0 & 0 &
 (\ref{sol6}) \\ \hline \hline
\end{tabular}
\end{table}

\begin{table}
\caption{Same as in Table~\ref{table1} but with numerators $\ge 5$.}
\label{table2}
 \begin{tabular}{|c|c|c|c|c|c|c|c|c|c|}\hline
$\,\,\nu\,\,$& $\,\,k_1\,\,$& $ \,\,k_3\,\,$& $\,\,m_1\,\,$ &
$\,\,m_2\,\,$ &$ \,\,n\,\,$ & $\,\,S\,\,$ &$\,\, V\,\,$&$ \,\,M\,\,$
& Eq. No.\\ \hline\hline $5/6$ & 1 & 2 & 1 & 1 & 0 & 0 & 1/5 & 0 &
 (\ref{sol7a}-\ref{sol7b}) \\ \hline \hline $5/7$ & 1 & 2 & 0 & 1 & 1 & 0 & 3/5 & 0 &  (\ref{sol7a}-\ref{sol7b})
\\ \hline\hline $5/8$ & 1 & 1 & 1 & 2 & 1 & 0 & 1/5 & 0 &  (\ref{sol7a}-\ref{sol7b}) \\
\hline $5/8$ & 1 & 2 & 1 & 0 & 1 & 0 & 1/2 & 0 &   (\ref{sol7a}-\ref{sol7b}) \\
\hline\hline
$5/9$ & 2 & 2 & 1 & 4 & 0 & 0 & 1/5 & 0 &  (\ref{sol7a}-\ref{sol7b}) \\ \hline
\hline $5/12$ & 1 & 2 & 2 & 3 & 2 & 0 & 3/5 & 0 &  (\ref{sol7a}-\ref{sol7b}) \\ \hline
$5/12$ & 2 & 2 & 0 & 3 & 2 & 0 & 3/5 & 0 &  (\ref{sol7a}-\ref{sol7b}) \\ \hline \hline
 $5/13$ &2 & 2 & 1 & 2 & 2 & 0 & 1/5 & 0 &   (\ref{sol7a}-\ref{sol7b}) \\ \hline\hline $5/18$ & 2 &
2 & 3 & 4 & 3 & 0 &1/5 & 0 &  (\ref{sol7a}-\ref{sol7b}) \\ \hline \hline $7/10$ & 1 & 2 &
0 & 3 & 1 & 0 & 5/7 & 0 &  (\ref{sol7a}-\ref{sol7b}) \\ \hline \hline $7/12$ & 1 & 2 & 1 &
2 & 1 & 0 & 3/7 & 0 &  (\ref{sol7a}-\ref{sol7b}) \\ \hline $7/12$ & 1 & 2 & 3 & 3 & 0 &
-- & 1/7 & -- &   (\ref{sol10}) \\ \hline $7/12$ & 2 & 2 & 1 & 3 & 0 & 0 & 1/7
& 0 &   (\ref{sol7a}-\ref{sol7b}) \\ \hline \hline $7/13$ & 1 & 2 & 2 & 1 & 1 & 0 & 1/7 &
0 &  (\ref{sol7a}-\ref{sol7b}) \\ \hline $7/13$ & 2 & 2 & 0 & 1 & 1 & 0 & 1/7 & 0 &   (\ref{sol7a}-\ref{sol7b})
\\ \hline\hline $7/19$ & 2 & 2 & 1 & 4 & 2 & 0 & 3/7 & 0 &  (\ref{sol7a}-\ref{sol7b}) \\
\hline \hline $8/9$ & 1 & 2 & 0 & 4 & 0 & 0 & 1/2 & 0 &  (\ref{sol7a}-\ref{sol7b}) \\
\hline \hline $8/11$ & 1 & 1 & 0 & 2 & 1 & 0 & 1/2 & 0 &  (\ref{sol7a}-\ref{sol7b}) \\
\hline \hline $8/19$ & 1 & 2 & 2 & 2 & 2 & 0 & 1/2 & 0 &   (\ref{sol7a}-\ref{sol7b}) \\
\hline $8/19$ & 2 & 2 & 0 & 2 & 2 & 0 & 1/2 & 0 &    (\ref{sol7a}-\ref{sol7b}) \\ \hline
\hline

\end{tabular}

\end{table}

\end{document}